\documentclass[prl,twocolumn,showpacs,preprintnumbers,amsmath,amssymb]{revtex4}

\usepackage[dvips]{graphics}
\usepackage{tikz}

\begin{document}

\title{Spacetime as a Tightly Bound Quantum Crystal}

\author{Vlatko Vedral}
\affiliation{Clarendon Laboratory, University of Oxford, Parks Road, Oxford OX1 3PU, United Kingdom and\\Centre for Quantum Technologies, National University of Singapore, 3 Science Drive 2, Singapore 117543 and\\
Department of Physics, National University of Singapore, 2 Science Drive 3, Singapore 117542}

\date{\today}

\begin{abstract}
\noindent We review how reparametrization of space and time, namely the procedure where both are made to depend on yet another parameter, can be used to formulate quantum physics in a way that is naturally conducive to relativity. This leads us to a second quantised formulation of quantum dynamics in which different points of spacetime represent different modes. We speculate on the fact that our formulation can be used to model dynamics in spacetime the same way that one models propagation of an electron through a crystal lattice in solid state physics. We comment on the implications of this for the notion of mode entanglement as well as for the fully relativistic Page-Wootters formulation of the wavefunction of the Universe. 
\end{abstract}


\maketitle                           

It is well known that the ordinary, non-relativistic, quantum mechanics treats space and time differently. 
The state of the system, which is defined for all space and is specified at a particular instant of time, is first given at a particular time and then evolved in time using the Schr\"odinger equation. If the system is multipartite, then each of the subsystems inhabits its own Hilbert Space but the dynamics is handled in the same way. Upon second quantisation, each point in space becomes a mode and is assigned its own creation and annihilation operators for the particles under consideration, while time still remains a parameter. The state of the system is then given as the state of all the modes at one time and is again evolved in time using the second-quantised Schr\"odinger equation. 

We have recently been investigating the possibility of writing the state of a quantum system to include different times on a equal footing with space \cite{Fitzsimons, Marletto}. When different times are assigned different Hilbert spaces (or when they become different modes upon second quantisation \cite{Zhang}), the resulting state of the system - called a pseudo-denisty matrix - may have negative eigenvalues. This presents a difficulty if we are to interpret them as probabilities as is the case in standard quantum mechanics. The reason for this non-positivity of the pseudo-density matrix is the fact that temporal correlations reflect non-commutativity of different observables, while, when it comes to different spatial points, all observations always commute. Of course, time is not just another dimension like space, and the differences are expected to appear in our description, however, one wanders if time and space could still be ``packaged" together to reflect the relativistically imposed symmetries.  

Here we present a way to treat both space and time as modes, while preserving the positive character of the overall spacetime state. This will be achieved in two steps. First, we will rewrite the standard single particle Schr\"odinger equation in a reparametrised way so that both space and time become functions of another parameter. The physical meaning of this parameter will deliberately be left open in this work, although see \cite{Nambu} where the parameter is interpreted as the proper time of the particle. The resulting equation (which can be made relativistically compliant as shown below) is then second quantised to obtain modes as pertaining to spacetime points, such that the assigned creation and annihilation operators now create and annihilate particles at different spacetime points (instead of doing so across all space and at a given instance of time). We then interpret this as a one dimensional chain of spacetime points, where this single dimension corresponds to the newly introduced parameter, and the state of our system is defined on it globally. The state then evolves, in this new parameter, according to the second quantized reparametrised Schr\"odinger equation. This is mathematically analogous to an electron hopping between different spatial sites in the Hubbard model, it is just that the hopping takes place in the ``fifth dimension" and the sites themselves are now spacetime points (see e.g. \cite{Ichiyanagi} for the treatement of quantum tunneling using the reparametrised formalism).  

We first formally present the above-outlined construction and then proceed to discuss how the resulting model could be used to explore the territory going beyond the standard quantum mechanics and possibly incorporate the quantum effects of the gravitational field (if there are any such effects). The main aim of this work is to present the formalism and point at some conceptual issues. We will finally show that the reparametrised Schr\"odinger equation can actually be presented in a way that dispenses with the need for dynamics. 

We start with the standard Schr\"odinger equation for a single particle: 
\begin{equation}
i\hbar \frac{\partial}{\partial t} \Psi(x,t) = H(x) \Psi(x,t) \; ,
\end{equation}
where the system is assumed to be closed, the Hamiltonian time-independent, and $x$ could represent all three spatial dimensions. We now express both space and time as a function of another parameter $\tau$, such that $x=x(\tau)$ and $t=t(\tau)$. As far as the new parameter is concerned, the original space and time are treated equally (though of course, they could transform differently as is the case in relativity). This leads to the following equation in $\tau$ \cite{Nambu,Ichiyanagi,Deriglazov}: 
\begin{equation}
i\hbar \frac{\partial}{\partial \tau} \Psi(x,t;\tau) = \left(-i\hbar\frac{\partial}{\partial t} + H(x)\right) \Psi(x,t;\tau) \; ,
\end{equation}  
which has the same structure as the original equation, but with the new Hamiltonian defined as $\hat E = -i\hbar\frac{\partial}{\partial t} + H(x)$. This equation is equivalent to the original non-relativistic equation, it is just that we are taking a perspective from the fifth dimension $\tau$, or from a world-line traced in spacetime. In that sense, the wavefunction in the reparametrised equation evolves from the initial spacetime point $x_0,t_0$ to the final spacetime point $x_0,t_0$ along a trajectory specified by the parameter $\tau$. 

The above equation can be solved using the standard technique of separation of variables. First we assume that
\begin{equation}
\Psi(x,t;\tau) = \psi(x,t) e^{-iE\tau/\hbar}
\end{equation}
which, when substituted into the reparametrized equation, yields: 
\begin{equation}
E\psi(x,t) =  \left(-i\hbar\frac{\partial}{\partial t} + H(x)\right) \psi(x,t) \; .
\end{equation}
Since $E$ is just a constant with units of energy that could be absorbed into the Hamiltonian $H(x)\rightarrow H(x)+E$, this implies that the wavefunction $\psi(x,t)$ obeys the usual Schr\"odinger equation. The value of $E$ does not have any direct physical meaning unless it is possible to interfere branches with different values of $E$. From the present perspective, the advantage of introducing $E$, along with $\tau$, is simply to be able to treat $x$ and $t$ on an equal footing. 

There are no obstacles to making the reparametrised equation relativistically complient (so that it is equivalent to, say, the Klein-Gordon equation for a free particle) \cite{Fanchi}:
\begin{equation}
i\hbar \frac{\partial}{\partial \tau} \Psi(x,t;\tau) = \frac{1}{2m}\left(\nabla^2 -\frac{\partial^2}{c^2\partial t^2}\right) \Psi(x,t;\tau) \; , 
\label{relativ}
\end{equation} 
where the form of the Hamiltonian is guided by the fact that it should have the units of energy, $E=p^{\mu}p_{\mu}/2m$ (from here on we will use the letter $H$ for the Hamiltonian even when reparametrised). Here  $p^{\mu}p_{\mu}=p^{0}p_{0} - {\bf p p}$ is the square of the relativistic $4$-momentum.  This equation is constructed so that $\psi (x,t)$ governs a single relativistic free particle, however, the same method could be applied to reproduce any other desired dynamics. It can also be made to include many particles as well as superpositions of states with different numbers of particles (as we will see below).

The relativistic invariance is transparent as the Hamiltonian is itself invariant and the state is a scalar $\Psi'(x',t';\tau')=\Psi(x,t;\tau)$. Here $x',t'$ are related to $x,t$ through a Lorentz transformation and $\tau'=\tau$. The parameter $\tau$ does not transform, since it is just an arbitrary label. Its purpose is only to encode the dynamics in what mathematically represents a fifth dimension. (If, on the other hand, $\tau$ is interpreted as a proper time \cite{Nambu}, it is then anyway a relativistic invariant). 

Before we proceed with the second quantisation of the equation (\ref{relativ}), which would enable us to include particle creation and annihilation proceeses, we first discuss its solutions. Let us define the eigenstates of the square of the $4$-momentum operator:
\begin{equation}
p^{\mu}p_{\mu} \psi_q (x,t) = q^2 \psi_q (x,t) \; ,
\end{equation} 
such that they satisfy the following orthogonality condition:
\begin{equation}
\int \int \psi^*_q (x,t) \psi_r (x,t) dxdt = \delta_{qr} \; ,
\end{equation} 
where $\delta_{qr}$ is the usual Kronecker delta function. In analogy with the non-relativistic case,
the general solution of the reparametrised Klein-Gordon equation can then be written as:
\begin{equation}
\Psi(x,t;\tau) = \sum_q A_q \psi_q (x,t) e^{iq^2\tau/2m}
\label{state}
\end{equation} 
where $A_q$ is just the amplitude corresponding the $q$-th eigenstate whose phase factor is $e^{iq^2\tau/2m^2}$. 

What is the physical meaning of this wavefunction? It is simply constructed so
that $\psi (x,t)$ satisfies the Klein-Gordon equation which means that $\omega^2 - k^2/c^2 = q^2$. Classically, $p^{\mu}p_{\mu}= E^2/c^2 -p^2 = m_0^2 c^2$. This implies that the value $q$ can be interpreted as the rest mass of the particle (times the speed of light). The above wavefunction is a superposition of different values of the mass, which may or may not be physically accessible (just like the overall energy may or may not be an observable in the non-relativistic case). 

In order to explore this further, let us compute the expected value of the $p^{\mu}p_{\mu}$ operator in the above state in eq.(\ref{state}):
\begin{eqnarray}
\langle p^{\mu}p_{\mu}\rangle & = & \int \int \Psi^* (x,t;\tau) \left(\nabla^2 -\frac{\partial^2}{c^2\partial t^2}\right)\Psi (x,t;\tau) dxdt  \nonumber \\
& = & \sum_q |A_q|^2 q^2 = \sum_{\omega,k} |A(\omega,k)|^2  (\omega^2 - k^2/c^2) \; .
\end{eqnarray} 
Quantumly, the classical relationship holds on average which implies that the mass could be thought of as a quantum operator. In fact, in addition to the position and momentum being conjugate, here the time and frequency (represented by the operator $-i\hbar \partial/\partial t$) are also conjugate. The quantum-classical correspondence then follows along the lines of the Ehrenfest theorem. If the parameter $\tau$ is interpreted as the proper time of the particle \cite{Nambu}, the same average relationship exists for the relativisitic equation $c^2 t^2 - x^2 = c^2 \tau^2$. This means that the classical relativisitc equations only hold in the classical limit of quantum mechanics and are otherwise subject to the quantum fluctuations due to the Heisenberg Uncertainty relations. This would also imply that there are some quantum corrections to the classical causal structure of the Minkowski spacetime but we leave this topic unexplored here \cite{Fanchi}.

We can also proceed with defining the spacetime propagator as 
\begin{equation}
A(x, t;x_0,t_0) = \langle x, t| e^{-i\frac{p^{\mu}p_{\mu}}{2m}\tau} |x_0,t_0\rangle\; .
\end{equation}
This is straightforward to compute:
\begin{eqnarray}
& & \int d^3 k e^{k(x - x_0)}e^{-\frac{ik^2}{2m\hbar}\tau}\int d^3 k e^{\omega (t - t_0)}e^{-\frac{i\omega^2}{2mc^2\hbar}\tau} = \nonumber \\
& & \left( \frac{m\hbar}{2\pi i \tau}\right)^{3/2}e^{-\frac{im(x-x_0)^2}{2\tau}} \times \left( \frac{mc^2\hbar}{2\pi i \tau}\right)^{1/2}e^{-\frac{im(t-t_0)^2}{2\tau}} \nonumber\; ,
\end{eqnarray}
and is, possibly not surprisignly, the product of the spatial and the temporal propagators. There is no novelty here other than offereing an alternative way of treating the reparametrised dynamics. 

We now focus on the main topic, namely obtaining the second quantised version of the reparametrised Schr\"odinger equation. First we promote the wavefunction into an operator: $\hat\Psi(x,t;\tau) = \hat \psi(x,t) e^{-iE\tau/\hbar}$ which satisfies the equation $H(x,t)\psi(x,t) = E\psi(x,t)$. We can expand  $\hat \psi(x,t)=\sum_n \hat a_n(t)\psi_n (x)$ (in the Heisenberg picture) in terms of spatial modes $\psi_n (x)$ which form an orthonoral basis such that $\int \psi^*_n (x) \psi_m (x) d^3 x = \delta_{nm}$.  The second quantised Hamiltonian is now given by
\begin{eqnarray}
\tilde H & = & \int d^3 x dt \hat\psi^{\dagger}(x,t) H(x,t)  \hat\psi (x,t)\nonumber \\
& = & E \sum_n a^{\dagger}_n a_n
\end{eqnarray}
which is simply the energy times the average number of particles with that energy (and we have dropped the hats). 

To make the analysis more concrete and to be able to explore some further aspects of the reparmatrized formalism, we use the plane-wave expansion, such that $\psi (x,t) = e^{i(\omega t - k x)}$. Then 
\begin{equation}
\Psi(x,t;\tau) = \int \int A(k,\omega) e^{i(\omega t - k x)} e^{i\hbar(k^2 - \omega^2/c^2)\tau/2m} d^3 k d\omega \nonumber \; ,
\end{equation} 
where the state is always normalised so that $\int |\Psi(x,t;\tau)|^2 dxdt=1$ for every $\tau$. This simply represents the fact that the particle has to exist at every point of the parametrisation. The quantity $|\Psi(x,t;\tau)|^2dxdt$ is therefore the probability that the particle occupies an infinitesimal volume surrounding the spacetime point $(x,t)$ and it is by definition a positive quantity, just like in standard quantum theory. This clearly demonstrates that it is possible to obtain a wavefunction that represents a state spreading across both space and time. 

Interestingly, the Fourier transform (in $\tau$) of the above state is given by \cite{Pavsic}
\begin{eqnarray}
\Phi (x, t; \mu) & = & \int_{-\infty}^{+\infty} d\tau \Psi(x,t;\tau) e^{i\mu\tau} \nonumber \\
& = & \int d^3 k d\omega A(k,\omega) e^{i(\omega t - k x)} \delta (k^2 - \omega^2/c^2 + \mu) \nonumber
\end{eqnarray}
which is simply a state with a well defined mass $k^2 = \omega^2/c^2 - \mu$ and so $\mu = q^2$. The Fourier transform therefore serves as a way of ensuring the sharpness of the mass. 
The above shows us how a relativistically invarient state can be written with the additional on the mass shell condition.  

The second quantization can now be performed with these states (see also \cite{Franchi-field}). The wavefunction $\Psi(x,t;\tau)$ becomes a quantum field expanded in terms of (now the operator) $\psi (x,t)$ which is to be interpreted as the annihilation operator for a particle at the spacetime point $(x,t)$. Also,  $\psi (x,t)=\int e^{i(\omega t - k x)} a(k,\omega) d^3 k d\omega$ where $a(k,\omega)$ annihilates the particle with momentum $k$ and energy $\omega$. The second-quantised Hamiltonian is as before given by
\begin{equation}
H = \int d^3 x dt \psi^{\dagger} (x,t) \frac{1}{2m}\left(\nabla^2 -\frac{\partial^2}{c^2\partial t^2}\right)  \psi (x,t)
\end{equation}
and it can be computed to be:
\begin{equation}
H = \int \frac{\hbar^2(\omega^2/c^2-k^2)}{2m} a^{\dagger}(k,\omega)a(k,\omega) d^3kd\omega 
\end{equation}
which represents a free field whose energy is $\hbar^2(\omega^2/c^2-k^2)/2m = mc^2/2$ and $a^{\dagger}(k,\omega)a(k,\omega)$ is the number operator corresponding to that energy. Note that the integral above includes both positive as well as negative $\omega$, the latter being linked with the existence of anti-particles (as we will see below). This Hamiltonian is not the same as the conventional one, $\sum_k \hbar \omega_k a^{\dagger}(k)a(k)$, because we are creating a particle in both $k$ and $\omega$ and their relationship is fixed through the particle's rest mass. 

Microcausality is ensured by the commutator $[\Psi(x,t;\tau), \Psi^\dagger(x',t';\tau')]$ 
which is equal to $\delta (x-x')\delta (t-t')$ for $\tau = \tau'$. Integrating over $d(\tau-\tau') e^{i\mu (\tau-\tau')}$ leads us to \cite{Pavsic-2}:
\begin{equation}
\Delta (x-x',t-t')=\int d^3 k d\omega \delta (p^2 - \mu) e^{i(\omega (t-t') - k (x-x'))} \nonumber \; . 
\end{equation}
This expression is Lorentz invariant and it enforces the mass shell constraint. The second quantisation can be made to obey micro-causality just like any other standard approach to quantum field theory \cite{Nambu}. This is only natural given that we can make a simple connection with the standard quantum field theory. Namely, the field state $\phi (x,t; \mu)=\int d^3 k d\omega A(k,\omega) e^{i(\omega t - k x)} \delta (k^2 - \omega^2/c^2 + \mu)$ can be written as 
\begin{equation}
\phi (x,t; \mu)=\int_{\omega \geq 0} \frac{d^3 k}{2\omega (k)} a(k) e^{i(\omega t - k x)} + c^* (-k)  e^{-i(\omega t - k x)}
\end{equation}
by using a simple expression involving the delta function and positive frequencies only.  This is just the standard quantum field theoretic decomposition in terms of the particle annihilation and anti-particle creation operators ($c^{\dagger}$) \cite{Fanchi,Schweber}. 

We note that because the second quantized formulation allows for states with many particles (even superpositions of such states), it is hard to interpret $\tau$ as a proper time of these individual particles. However, $\tau$ could still represent a world-line of a single observer who is describing the dynamics of the quantum field. This is very much in the spirit of the relativistic ``block" universe picture of dynamics, where the observer moves across a static spacetime (where all the events are given once and forever), each $\tau$ representing a different moment of ``now" for that observer. 

This, of course, is not to suggest that observers have any special status in quantum physics (any more than they do in relativity). An observer could simply be another quantum system \cite{Vedral-observer}, such as an atomic clock and $\tau$ could be its proper time or, indeed, any other observable that is easily accessible and suitable for the task. It is also possible to have a superposition of observers, just like in ordinary quantum physics, each with its own proper time. Needless to say, a superposition of different $\tau$s could always be phrased with respect to yet another parameter $\sigma$ and so on, an infinitum, much as in quantum mechanics we can have an infinite chain of observers whose observations are all consistent with each other.  

The second quantization allows us to create and annihilate particles, which presents us with an intriguing opportunity to view the evolution of a quantum particle as a process of hopping between different points of spacetime. Namely, at some value of the parameter $\tau$ the particle is destroyed, while at the point $\tau+\delta \tau$ it is created. This seems a natural way to think about the reparametrised quantum dynamics. To utilise this picture fully, one could add an interaction to our free field. 

So far, spacetime is used only for parametrisation of the dynamics (so that any conventional problem in quantum mechanics could be addressed with this method), however, we can make it play a more active role. This is certainly warranted if we are to think of spacetime as a physical entity in its own right. The question then is how the spacetime couples to a moving particle (or to an evolving quantum field). We warn the reader that what follows is entirely speculative. To explore this possibility we make the assumption that the state of the particle can be localised at a discrete set of spacetime points (much as electrons are localised close to atomic nuclei in a typical crystal lattice). At present we will treat this spacetime crystal as a background potential (i.e. classically) although there is always a possibility to perform a fully quantum treatment as one does with various phononic models in the solid state (so that the particle can actually become entangled to spacetime - as in e.g. \cite{Vedral}). The exploration of the fully quantised model will be left for future work. At present, and for the sake of illustration, let us assume that the bound particle states are given by the (spacetime) Wannier functions of the form:
\begin{equation}
\psi_n (x,t) \propto e^{-\frac{x-x_n}{\sigma_x}}e^{-\frac{t-t_n}{\sigma_t}} \; ,
\end{equation}
where $(x_n,t_n)$ are the coordinates of the $n$-th site and $\sigma_x$ and  $\sigma_t$ are
the dispersions in space and time respectively. We proceed as one would in the solid state domain.
The second quantised tight-binding hopping amplitude (we assume that only the nearest neighbour interaction is relevant) is computed as:
\begin{eqnarray}
J & = & \int \psi^*_n (x,t) \left(\nabla^2 -\frac{\partial^2}{c^2\partial t^2}\right) \psi_{n+1} (x,t) dxdt\nonumber \\
& = & \frac{\hbar^2}{2m}\frac{\sigma^2_\tau}{\sigma^2_x\sigma^2_t} e^{-\frac{\Delta x \sigma_t-\Delta t\sigma_x}{\sigma_x\sigma_t}} \; ,
\end{eqnarray}
where $\Delta x = x_n - x_{n+1}$ and $\Delta t = t_n - t_{n+1}$. The second quantised Schr\"odinger equation is then given by:
\begin{equation}
J \sum_n (b_n b^{\dagger}_{n+1} +b^{\dagger}_n b_{n+1}) |\Psi (\tau)\rangle = i\hbar \frac{\partial}{\partial \tau} |\Psi (\tau) \rangle \; ,
\end{equation}
where we have omitted the term with the onsite energy (which could be assumed to be zero). The interpretation of this equation is straightforward. It describes the hopping between the $n$-th and $n+1$st spacetime sites, whose strength is governed by $J$.  The Hamiltonian can be exactly diagonalised (see any standard solid-state textbook, e.g. \cite{Son}) and we can proceed to define the effective mass as one normally does for the tight-binding model. It might be natural to assume that the spacetime discretisation
is on the order of Planck scales. The main question then, of course, is to investigate if our approach can lead to novel predictions regarding any effect of this on particle dynamics. 

Reformulating the concept of mode entanglement within the ``spacetime points--as--modes" picture also seems like a worthwhile venture. If a single particle is superposed across two such modes, then the state would be written as $|0\rangle_{x_1t_1}|1\rangle_{x_2t_2}+|1\rangle_{x_1t_1}|0\rangle_{x_2t_2}$. When $t_1=t_2$ and $x_1\neq x_2$ this state is the standard spatial mode single particle entangled state. However, when $t_1\neq t_2$ and $x_1 = x_2$, this state represents a purely temporally entangled state (cf. \cite{ent-time}), i.e. a particle that exists at the same spatial point but superposed across two times. A possible physical interpretation of this state is offered in \cite{Vedral-twin}, where it is show how to use the relativistic time dilation to create a superposition of a younger and older version of the same system (which in the second quantised notation corresponds to the state $|0\rangle_{x_1 t_1}|1\rangle_{x_1t_2}+|1\rangle_{x_1t_1}|0\rangle_{x_1t_2}$, i.e. spacially the system is at the same point, but temporally it is extended). Therefore, here the internal time of the system is in a superposition with respect to another, call it external time. This is another indication of the fact that the labels $t$ and $\tau$ are somewhat arbitrary, since the proper time would in this example be $t$ and $\tau$ would correspond to the extrnal clock that is used to prepare the requires temporal superposition.

The principle of reparametrisation invariance plays an important role in a number of areas such as the constrained quantisation \cite{Dirac} as well as in String Theory \cite{Zwiebach}. However, we would like to apply reparametrisation to a problem with which it seems to be even more intimately linked, namely the Page-Wootters \cite{Page} ``timeless" formulation of quantum physics. Our analysis points to a simple way to make the Page-Wootters construction compliant with relativity. To do so, we assume that the state of the universe, which in our case will be a particle and a clock, is given by
\begin{equation}
|\Psi\rangle = \sum_{\tau} |\psi (x,t;\tau)\rangle |\tau\rangle \; , 
\end{equation}
where now the parameter $\tau$ plays the role of the ``clock" \cite{Marletto-Vedral}. We furthermore postulate that this state is an eigenstate of the total Hamiltonian $(H_s + H_c)|\Psi\rangle =q^2/2m |\Psi\rangle$, where $H_s$ is the Klein-Gordon Hamiltonian of the system and $e^{-iH_c \Delta \tau} |\tau\rangle=|\tau+\Delta \tau \rangle$. Relative to the states of the clock, the system now undergoes the evolution according to the Klein-Gordon equation. We have used a discretised parameter $\tau$ just as in our tight-binding model, but this is done without any loss of generality since $\tau$ can be made arbitrarily small. The Page-Wootters formulation of relativistic dynamics is simply the integral version of the reparametrised relativistic differential equations. 

The physical interpretation of the clock is again not crucial, but it could certainly represent an internal degree of freedom of the particle that serves as a measurer of the proper time. This reinforces the idea that dynamics could be seen as a consequence of the underlying symmetries and once these symmetries are specified (in the form of the relativistic spacetime correlations) the dynamics follows (or, rather, dynamics is just another way of expressing the same correlations that could alternatively also be parametrised in a ``timeless" fashion). 

Finally, in quantum field theory we normally assume that field operators commute at spacelike separated points. However, it could be that, because of the extended nature of the Wannier spacetime functions, the operators that are centred at spacelike separated spacetime points, still have a non-zero overlap with one another \cite{fuzzy} (the overlap would depend on the ratio between $\sigma_\tau$ and $\Delta \tau$, since if $\sigma_\tau \geq \Delta \tau$ then neighbouring points would have a significant overlap). Even though this scenario may still respect the fact that no singnal can travel faster than the speed of light (in whichever incarnation this is to be understood), it may have to change the way that we impose commutation relations on different field components. A discretisation of this type may help us deal with various infinities in quantum field theory, though its physical origin would have to be understood (see also \cite{David}).  This and related questions certainly appear very much worth exploring in the future.

\textit{Acknowledgments}: The author acknowledges extensive discussions  on many topics related to this work with Aditya Iyer, Eduardo Dias, Chiara Marletto and Nicetu Tibau Vidal. This research is supported by the National Research Foundation and the Ministry of Education in Singapore and administered by Centre for Quantum Technologies, National University of Singapore.  This publication was made possible through the support of the ID 61466 grant from the John Templeton Foundation, as part of the The Quantum Information Structure of Spacetime (QISS) Project (qiss.fr). The opinions expressed in this publication are those of the authors and do not necessarily reflect the views of the John Templeton Foundation.

\end{document}